\documentclass{JHEP}
\usepackage{amsmath,amssymb}
\newcommand{\Tr}{{\rm Tr\,}}
\newcommand{\be}{\begin{equation}}
\newcommand{\ee}{\end{equation}}
\newcommand{\eq}[1]{(\ref{#1})}
\newcommand{\Dslash}{\ensuremath \raisebox{0.025cm}{\slash}\hspace{-0.32cm} D}
\def\nn{\nonumber}
\def\bea{\begin{eqnarray}}
\def\eea{\end{eqnarray}}
\title{One-Loop Effect of  Null-Like  Cosmology's Holographic Dual Super-Yang-Mills}

\author{Feng-Li Lin and Dan Tomino\\
Department of Physics, National Taiwan Normal University\\
Taipei, Taiwan 116\\
E-mail: \email{linfengli@phy.ntnu.edu.tw},\email{dan@phy.ntnu.edu.tw}}

\abstract{We calculate the 1-loop effect in super-Yang-Mills which
preserves $1/4$-supersymmetries and is holographically dual to the
null-like cosmology with a big-bang singularity. Though the bosonic and
fermionic spectra do not agree precisely, we do obtain vanishing
1-loop vacuum energy for generic warped plane-wave type backgrounds
with a big-bang singularity. Moreover, we find that the cosmological ``constant"
contributed either by bosons or fermions is time-dependent. 
The issues about the particle production of some background and about the UV structure are also commented. We argue that the effective higher derivative interactions are suppressed as long as  the Fourier transform of the time-dependent coupling is UV-finite. Our result holds for scalar configurations that are BPS but with arbitrary time-dependence. This suggests the existence of non-renormalization
theorem for such a new class of time-dependent theories.  Altogether, it implies that such a super-Yang-Mills is scale-invariant, and that its dual bulk quantum gravity
might behave regularly near the big bang.}


\begin{document}

%%%%%%%%%%%%%%%%%%%%%%%%%%%%%%%%%%%%%%%%%%%%%%%%%%%%%%%%%%%%%%%%%%%%%%%%%%%%
\section{Introduction}
%%%%%%%%%%%%%%%%%%%%%%%%%%%%%%%%%%%%%%%%%%%%%%%%%%%%%%%%%%%%%%%%%%%%%%%%%%%%
 An important issue in quantum gravity is how
the big bang singularity is resolved. However, the divergent
gravitational coupling due to the singular curvature near the big
bang invalidates the perturbative and semi-classical approach.
Therefore, non-perturbative approach to tackle this problem is
needed. Among the non-perturbative approaches, a less direct one is to
study the dual weakly-coupled field theory of the strongly-coupled
gravity theory. In string theory, two schemes of this strong/weak
S-duality are known. One is the BFSS matrix model and another is the
AdS/CFT correspondence. Both are of the type of gravity/field theory
dualities. Recently, these two schemes have been generalized to
supersymmetric null-like cosmological backgrounds. In both cases,
the dual field theories have time-dependent coupling constant. For
the matrix big bang, see \cite{mbb}, \cite{craps-r} and the
references therein; for the null-like time-dependent AdS/CFT, see
\cite{ch,lw,nullh}. For other recent related papers , see \cite{recent}.

 The dual field theory can be thought of as a probe in the bulk spacetime.
Even at tree level, it seems that the big-bang singularity is
resolved due to the emergence of the new non-perturbative degrees of
freedoms such as the branes \cite{mbb} or the divergence of the
holographic c functions, as studied in \cite{lw} for a toy example.
Despite that, it is not clear how one can translate the result back
to gravity and understand the resolution of the singularity in
the strong gravity regime. Moreover, one is not sure if the
resolution of the singularity persists even at the loop level of the
dual field theory. In both cases, the backgrounds preserve
fractional supersymmetry, and one may expect better UV behavior due
to the fermion-boson cancellation. However, in the matrix big bang,
the residual supersymmetry is completely spontaneously broken in the
dual matrix model description \cite{mbb}, so one would expect the
nontrivial loop effect. These issues have been examined in
\cite{1-loop-1} and \cite{1-loop-2}. In \cite{1-loop-1} the authors
studied the matrix quantum mechanics with time-dependent coupling
and found that the 1-loop effect is trivial for the static
configuration, but is nontrivial for the non-static one. On the
other hand, the authors in \cite{1-loop-2} studied the matrix string
compactified on the Milne orbifold and found non-trivial late time behavior of the effective potential even for the static scalar background
configuration, which implies the presence of the big-bang.  The possible reason for the difference between the
results in \cite{1-loop-1} and \cite{1-loop-2} may be due to the
compactness of the Milne orbifold in \cite{1-loop-2} rendering
different spectra from \cite{1-loop-1}.

  As for null-like time-dependent AdS/CFT, the dual Super-Yang-Mills(SYM) theory
preserves 1/4-supersymmetry as explicitly shown in \cite{ch}. So one
would expect trivial loop effect, especially the zero vacuum energy
for the chosen scalar vev preserving the 1/4-supersymmetry, i.e. BPS
scalar configuration. However, since the dual theory is living on
the time-dependent cosmological background with time-dependent
coupling and is not a conventional quantum field theory, it is not
clear if the fermion-boson cancellation in this case is complete or
not. Indeed, we find that the fermionic and bosonic spectra differ
by a pure imaginary term though their contributions to the 1-loop vacuum
energy is completely canceled. This suggests that the scale-invariance \footnote{Though the conformal invariance of {\cal N}=4 SYM is broken, as shown in \cite{ch}, there is still a residual scale-invariance that guarantees the power-law structure of the SYM correlators.} of the dual SYM is preserved up to 1-loop. For completeness, one also
needs vanishing beta function of gauge coupling to guarantee the
scale invariance. We will discuss this issue and argue that the beta function is zero if the time-dependent effective coupling is UV-finite. Altogether, Our result implies the existence of
the non-renormalization theorem for such a time-dependent theory. This also  supports  the conjecture in \cite{ch,conformal}, in which they found the
power-law behavior of the holographic correlators. Besides, we find that some background considered in this paper will induce particle production, in contrast to the null result for a specific case considered in \cite{conformal}.

  Before closing our remarks, it deserves mentioning that the quantum field
theory in plane-wave background was discussed long time
ago \footnote{We thank C. Bachas for pointing this out to
us.}, and the 1-loop triviality was noted \cite{deser,gibbons}
even for non-supersymmetric theory. The extension to string theory
in plane-wave background was also studied in \cite{horowitz} for
closed string, and recently in \cite{bachas,hikida} for open
string. In this context our work can be seen as an extension of the
above works to the warped plane-wave background containing a
big-bang, and we find that the 1-loop triviality no longer holds
unless supersymmetry is imposed. Moreover, our result holds for
arbitrary time-dependent BPS scalar background vevs, and suggests the 1-loop triviality
of the open string partition function in the dual time-dependent AdS
space, which generalizes the result explicitly calculated in \cite{hikida}. This is
non-trivial in the sense that the perturbative string theory in AdS
space is not available yet.

  The paper is organized as follows: in the next section we
lay out the model by expanding the dual SYM action around some
nontrivial time-dependent BPS scalar background configuration which
represents two D3-branes separated transversely. In section 3, we
evaluate the 1-loop vacuum energy by explicitly solving the
eigenvalue problem for generic null-like background. In doing so we
have to introduce the non-trivial measure factor due to the warping
factor in defining the inner product of the eigenfunctions. This is
the key concept for the calculation in the warped plane-wave space.
We then find the exact boson-fermion cancellation in vacuum energy
even thought their spectra do not match precisely.  In section 4, we
solve for two specific backgrounds for pedagogical purpose.  In section 5 we will briefly comment on the subtle issues about the particle production and about the UV structure of the theory. We
conclude and discuss the implication of our result in the last
section. In Appendix  we lay down the details of integrating out the quantum fluctuations to arrive the effective
action.

\section{The model}

  The SYM studied in this paper is dual to the following
null-like time-dependent asymptotically $AdS_5\times S^5$ background
\cite{lw,ch,nullh}
\begin{eqnarray}\label{metric10}
ds^2_{einstein}&=&
e^{2\rho}a(u)^2(-2dudv+h(u,r,\vec{x})du^2+d\vec{x}^2_{(p-1)})
+d\rho^2+d\Omega^2_{(5)})\nn\\
e^{\phi}&=&e^{\phi(u)}\nn\\
F_{(5)}&=&4(e^{4\rho}a(u)^4du\wedge dv\wedge \cdots \wedge d\rho
+\omega_{(5)})
\end{eqnarray}
where the scale factor $a(u)$, the plane-wave profile $h(u,x,\rho)$
and the dilaton/axion $\phi(u)/\chi(u)$ are constrained only by a
single equation
\begin{eqnarray}\label{consap}\hspace{-7mm}
2((\partial_u\ln a)^2-\partial_u^2\ln a)
-\frac{1}{2}\vec{\nabla}^2h-e^{2\rho}a^2(2\partial_{\rho}
h+\frac{1}{2}\partial_{\rho}^2h)
=\frac{1}{2}(\partial_{u}\phi)^2+\frac{1}{2}e^{2\phi}
(\partial_u\chi)^2.
\end{eqnarray}

  The dual SYM is living on the following cosmological background
\be \label{metric4}
ds^2_4:=g_{\mu\nu}dx^{\mu}dx^{\nu}=a(u)^2[\;-2dudv+h(u,x^i)du^2+dx_2^2+dx_3^2\;]
\ee
with the time-dependent coupling constant $g_{YM}=e^{\phi(u)/2}$.
Note that the exact time dependence of the scale factor will depend
on the functional form of $\phi(u)$ and $h(u,\vec{x})$. In this
paper, we will consider a very generic background with
\be\label{generic1}
a=a(u),\qquad \phi=\phi(u), \qquad \chi=0, \qquad
h=h_0+h_1(u)+h_2(\vec{x})
\ee
constrained by \eq{consap}. We will see that we can exactly solve
the one-loop problem of the dual SYM  for such a kind of
background. Moreover, it was shown in \cite{ch,lw} that the
resulting bulk gravity and dual SYM preserve only a quarter of the
maximal supersymmetry due to the background deformation.

    Though the spacetime is short of curvature singularity,
as shown in \cite{ch,lw,nullh}, it is geodesic incomplete
at $a=0$. This can be seen by solving the geodesic equation and yields
\be
{du \over d\lambda}={1\over a^2}
\ee
where $\lambda$ is the affine parameter. The geodesic completeness
can be seen as a kind of big-bang. It is then interesting to ask
what is the 1-loop behavior of the dual SYM near the big-bang. The
result reflects the behavior of the bulk gravity near the big-bang
via the holographic principle. That is, the perturbative dual SYM
serves as the quantum gravity theory in the bulk.

 The formal action of the dual SYM can be obtained from the dimensional
reduction of SYM theory living on the background \eq{metric10},
which contains the bosonic and fermionic parts
\be\label{S0}
S=S_b+S_f.
\ee

 Explicitly the bosonic part is
\begin{eqnarray}
\hspace{-10mm} S_b&=&\frac{1}{2} \int d^4x \sqrt{-g} \; e^{-\phi(u)}
a(u)^{-4} \Tr\Bigg[
F_{uv}^2+F_{ui}F_{vi}+F_{vi}F_{ui}+hF_{vi}^2-F^2_{ij} \nonumber\\
&&+a(u)^2(2D_uX^mD_vX^m+h(D_vX^m)^2-(D_iX^m)^2)+\frac{a(u)^4}{2}[X^m,X^n]^2
\Bigg]
\end{eqnarray}
\begin{eqnarray}
F_{\mu\nu}&:=&\partial_{\mu}A_{\nu}-\partial_{\nu}A_{\mu}-i[A_{\mu},A_{\nu}],\\
D_{\mu}X^m&:=&\partial_{\mu}X^m-i[A_{\mu},X^m].
\end{eqnarray}
In the above, the range of the indices run as $i=2,3$ and
$m,n=4,\cdots,9$. Hereafter, for simplicity we may omit the indices
associated with scalar fields $X^m$. Note especially
\begin{eqnarray}
\sqrt{-g}\; e^{-\phi(u)} a(u)^{-4} =e^{-\phi(u)}.
\end{eqnarray}

 The fermionic part written in the compact 10-dimensional ${\cal N}=1$ spinor is
\begin{eqnarray}
S_f=\int dx^{4}\sqrt{-g}e^{-\phi(u)}&\Bigg[&
-\frac{i}{4}\bar{\Psi}\Gamma^{\alpha}\overleftrightarrow{\nabla}_{\alpha}\Psi
+\frac{1}{2}\bar{\Psi}(\Gamma^{\alpha}[A_{\alpha},\Psi]+\Gamma^{m}[X^m,\Psi])
\Bigg].
\end{eqnarray}
\begin{eqnarray}
\Gamma^{\alpha}\nabla_{\alpha}:=\Gamma^{\alpha}
e_{\alpha}^{\mu}(\partial_{\mu}-\frac{1}{4}\omega_{\mu \beta\gamma
}\Gamma^{\beta\gamma})
\end{eqnarray}
where we have used 10-dimensional Gamma matrices:
$\Gamma^{\alpha},\Gamma^m$, $\alpha=0,1,2,3$ and $m=4 ... 9$ for
convenience. Here $\alpha$ and $m$ are local Lorentz indices. The
veilbein $e_{\alpha}^{\mu}$ and the spin-connection $\omega_{\mu
\alpha\beta }$ for the metric \eq{metric4} will be given later.

  In the following we would like to evaluate the 1-loop effective action for the
following fluctuation in the $SU(2)$ sub-sector of the $SU(N)$
gauge group, i.e.,
\begin{eqnarray}\label{sback}
X^m \Rightarrow B^m +X^m_{a}\sigma^a,\qquad A^{\mu}\Rightarrow
A^{\mu}_{a}\sigma^a \qquad a=1,2,3
\end{eqnarray}
with the background scalar vev
\be\label{bvev}
B^m:={1\over2}\;b^m\sigma^3
\ee
where $\sigma^a$'s are Pauli matrices. This background configuration
can be seen as two D3-branes separated by a transverse distance
$2b^m$.

   Since we would like to calculate the effective action for the nontrivial
scalar background \eq{sback},  the background field
method \cite{Weinberg} will be used. Therefore we choose the background
field gauge
\begin{eqnarray}
&&{\cal G}:= -\partial_{v}A_{u} - \partial_{u}A_{v} -h
\partial_{v}A_{v} + \partial_{i}A_{i}  -ia^2({\cal B}X)=0
\\
&&{\cal B}X:=[B,X],
\end{eqnarray}
and choose the gauge fixing action added to \eq{S0}
\be
S_{g.f.}=-{i\over2}\Tr \int d^4x \;e^{\phi} \;{\cal G}^2.
\ee

 The corresponding Faddeev-Popov ghost's determinant due to
the standard gauge-fixing  procedure is
\begin{eqnarray}
\det_{gh}(- \partial_{u}D_{v}-\partial_{v}D_{u} -h\partial_{v}D_{v}
+\partial_{i}D_{i} -a^2{\cal B}^2-a^2[B,[X,\;]]).
\end{eqnarray}
Therefore 1-loop ghost determinant up to the quadratic order is
\begin{eqnarray}
\det_{gh}(-2 \partial_{u}\partial_{v}-h\partial_{v}\partial_{v}
+\partial_{i}\partial_{i} -a^2{\cal B}^2):=\det_{gh}(\Delta_0).
\end{eqnarray}
Note that using the background vev \eq{bvev}, it is easy to see
\be\label{beq1}
{\cal B}^2=b^mb_m(1-\delta_{a,3})
\ee
where $a$ is the $SU(2)$ index.

  We then expand the action $S_b+S_{g.f.}$ around the scalar
background \eq{sback} up to the quadratic order in the fluctuation
fields $A_{\mu}$ and $X^i$, and the result consists of
\begin{eqnarray}
&&S_A=\frac{1}{2}\Tr\int d^4x\; e^{-\phi} \Bigg[
-(\Delta_1A_{u})A_v-A_v(\Delta_1A_u)-hA_v\Delta_2A_v+A_i\Delta_1A_i+2(HA_v)A_i
\nonumber\\\label{SA1}
 &&\hspace{10mm}
-2ie^{\phi}\partial_{\mu}(e^{-\phi}a^4g^{\mu\nu})A_{\nu}[B^m,X^m]
-2iA_{\mu}[\partial^{\mu}B^m,X^m]
 \Bigg]
\end{eqnarray}
for gauge field fluctuation $A_{\mu}$, and
\begin{eqnarray} \label{SX1} S_{X}
&=&\frac{1}{2}\Tr\int d^4x \; e^{-\phi} a^4 \Bigg[
-\partial^{\mu}B^m\partial_{\mu}B^m-\frac{1}{2}[B^m,B^n]^2
-2\partial^{\mu}B^m\partial_{\mu}X^m-2[B^m,B^n][B^m,X^n]\nonumber \\
&&-\partial^{\mu}X^m\partial_{\mu}X^m-X^m{\cal B}^2X^m \Bigg]
\end{eqnarray}
for scalar fluctuation $X$. In the above we have defined
\begin{eqnarray}
\Delta_0&:=& \square-a^2{\cal B}^2\label{delta0}\\
\Delta_1&:=& \square +\phi'\partial_v-a^2{\cal B}^2\\
\Delta_2&:=& \square -2\frac{h'}{h}\partial_v+\frac{\nabla_ih}{h}\partial_i-a^2{\cal B}^2\\
H&:=&-\phi'\partial_i-\nabla_ih\partial_v
\end{eqnarray}
where $'$ is the derivative with respect to $u$,  and 
the 4-dimensional Laplacian in the co-moving frame is
\be
\square:=-2\partial_{u}\partial_{v}-h\partial_v\partial_v
+\partial_{i}\partial_{i} .
\ee
\\

 On the other hand, the fermionic action $S_f$ up to the quadratic
order yields the 1-loop determinant
\begin{eqnarray}\label{fermid}
\det_{\Psi}i \Gamma^0a\;\Dslash
\end{eqnarray}
where
\be
\Dslash=\Gamma^{\alpha}\nabla_{\alpha}
+\frac{1}{2}\Gamma^{\alpha}(\partial_{\mu}e^{\mu}_{\alpha})+\frac{1}{2}\Gamma^{\alpha}\nabla_{\alpha}\log
(\sqrt{-g}e^{-\phi})-i\Gamma^{m}{\cal B}^m
\ee
and ${\cal B}^m\Psi=[B^m,\Psi]$. To evaluate \eq{fermid}, some proper
re-scaling of the fermion field is needed (as mentioned later on).

 To further simplify the fermionic determinant, we need the veilbein and spin connection
for the metric \eq{metric4}. The veilbein are
\begin{eqnarray*}
&&\left(\begin{array}{cc} e^{+}_{u} &e^{+}_{v}  \\ e^{-}_{u}
&e^{-}_{v}
\end{array}\right)
= a\left(\begin{array}{cc}
 1&0  \\ \frac{h}{2} &-1
\end{array}\right),
\qquad \left(\begin{array}{cc} e^{u}_{+} &e^{u}_{-}  \\ e^{v}_{+}
&e^{v}_{-}
\end{array}\right)=
\frac{1}{a}\left(\begin{array}{cc} 1 &0  \\ \frac{h}{2} &-1
\end{array}\right),
\qquad e^{i}_{x_i}=a\\
\\
&&g_{\mu\nu}=e^{\alpha}_{\mu}\eta_{\alpha\beta} e^{\beta}_{\nu},
\qquad\qquad \eta= \left(\begin{array}{cccc} 0&1&& \\ 1&0&&\\ &&1& \\
&&&1
\end{array}\right),
\end{eqnarray*}
and the non-vanishing spin connection 1-forms are
\begin{eqnarray}
\omega_{+-}=-\frac{a'}{a}du,\quad
\omega_{+i}=\frac{1}{2}\partial_{x_i}h du-{a'\over a} dx^i, \quad
\end{eqnarray}
Using the above we can have
\begin{eqnarray}\label{Dslash1}
\Dslash =\Gamma^+ \frac{1}{a}\left[
\partial_u+\frac{h}{2}
\partial_v-2\frac{a'}{a}
+\frac{1}{2}\partial_u\log (\sqrt{-g}e^{-\phi})
\right]-\Gamma^-\frac{1}{a}
\partial_v+\Gamma^i\frac{1}{a}\partial_{x_i}-i\Gamma^m{\cal B}^m
\end{eqnarray}
where Gamma matrices with local Lorentz indices satisfy
$\{\Gamma^{A}, \Gamma^{B}\}=2\eta^{AB}$.\\
Later on we will use the above to evaluate its eigen-spectrum to
obtain the explicit form of \eq{fermid}.

\section{The 1-loop effective potential }

  Now we would like to evaluate the effective potential for the
scalar background
\begin{eqnarray}\label{b9}
B^{m=9}={1\over 2}\;b(u)\sigma_3, \qquad B^{m\ne 9}=0,
%\qquad h=\mbox{const.}
\end{eqnarray}
where $b(u)$ is a function of $u$. Interestingly, this configuration
is BPS though it is time-dependent. This can be seen from the
supersymmetry variation, which is
\be
\delta \Psi \propto F_{MN}\gamma^{MN} \epsilon_0
\ee
with
\be
\gamma^u \epsilon_0=0
\ee
as required by 1/4-supersymmetry of the bulk background geometry.
Therefore, even though $F_{u9}=\partial_u B_9 \ne 0$, we still have
$\delta \Psi=0$ such that the configuration \eq{b9} is BPS preserving the same 1/4-supersymmetry of the bulk geometry.

\subsection{Bosonic sector}

 Plug the background \eq{b9} into the quadratic action
$S_A+S_X$ and perform a succession of Gaussian integrations in
the order over $A_i$, $A_v$, $A_u$ and $X$, then we obtain the
vacuum energy $\Lambda_{b}$ for the bosonic sector. The detailed
steps of the above Gaussian integrations are shown in the Appendix, here
we just write down the final result\footnote{This time-dependent
Coleman-Weinberg type method is also used in \cite{1-loop-2} to
calculate the 1-loop potential for the matrix big bang.}
\begin{eqnarray}
-\Lambda_b&=&3i\Tr\log i\Delta_3+i\Tr \log i\Delta_1
+\frac{1}{2}i\Tr\log i(H\frac{1}{\Delta_1}H-h\Delta_2)\nonumber\\
&&+\frac{1}{2}i\Tr \log\left( \Delta_1
\frac{-i}{H\frac{1}{\Delta_1}H-h\Delta_2}\Delta_1\right)
-i\Tr \log \Delta_0  \label{veb1} \nonumber\\
&&-\delta_{a,3}\Tr\int d^4x \frac{1}{2}e^{-\phi}a^2b'^2
\left(\overrightarrow{\partial_v}\frac{1}{\Delta_3}\overleftarrow{\partial_v}\right)
\\ \label{veb2}
&=&3i\Tr \log i\Delta_3+i\Tr \log i\Delta_1
+i\Tr \log \Delta_1-i\Tr \log \Delta_0 \nonumber\\
&&-\delta_{a,3}\Tr\int d^4x \frac{1}{2}e^{-\phi}a^2b'^2
\left(\overrightarrow{\partial_v}\frac{1}{\Delta_3}\overleftarrow{\partial_v}\right)
\end{eqnarray}
where
\be
\Delta_3:=\square +\left(\phi'-2\frac{a'}{a}\right)\partial_v
-a^2{\cal B}^2.
\ee
Note that we have formally used the factor
$i(H\frac{1}{\Delta_1}H-h\Delta_2)$ in the 3rd $\log$ of \eq{veb1}
to cancel its inverse in the 4th $\log$. This formal operation can
be justified if we can find an orthonormal complete eigen-basis for
the operator $\Delta_1$, and we will see that it is indeed the case.
We also note that the ghost contribution $i\Tr \log \Delta_0$
cancels the contribution $i\Tr \log \Delta_1$ which come from two of
the gauge fields in flat space.

 In performing the Gaussian integration over the
fluctuation fields, we find that the kinetic terms are not canonical
in terms of the flat space Laplacian, i.e. $\Delta_l$. Instead, they
are dressed by some Weyl factor, i.e. $G_l \Delta_l$. From each step
of the Gaussian integrations over the fluctuation fields explicitly
shown in Appendix, we can read off $G_l$ as following
\be\label{Gs1}
G_0:=1, \qquad G_1:=e^{-\phi}, \qquad G_3:=e^{-\phi} a^2.
\ee
In the formal expressions of \eq{veb1} and \eq{veb2} we have omitted
the factor $G_l$. However, in evaluating the determinants we should
find the orthonormal and complete set of eigenfunctions of
$\Delta_l$ with respect to the ``measure" $G_l$.

  To proceed, we need to solve the corresponding
eigenvalue equation in the following form
\be\label{eigen1}
\Delta_l \Phi_{l,Q_E}=E_l\Phi_{l,Q_E}
\ee
for $l=0,1,3$, in which $Q_E$ labels the momentum associated with $E_l$. 
With the help of \eq{beq1} we have
\be
\Delta_l=\square+\gamma_l\partial_v-F(u)(1-\delta_{a,3})
\ee
with $F(u)=a(u)^2b(u)^2$ and
\be\label{gammal}
\gamma_0:=0,\;\;\gamma_1:=\phi',\;\;  \gamma_3:=\phi'-2(\log a)'
\ee
which correlate with $G_l$'s given in \eq{Gs1} by
$\gamma_l=-G_l'/G_l$.

One can expand the fluctuation fields in terms of the corresponding
$\Phi_{l,Q_E}$. Note that $\Phi_{l,Q_E}$'s are orthonormal and
complete with respect to the measure $G_l$, i.e.,
\be\label{or1}
\int dY G_l(Y) \Phi^*_{l,Q_E}(Y) \Phi_{l,Q_E'}(Y)=\delta(Q_E-Q_E'),
\ee
and
\be\label{co1}
\int dQ_E \; G_l (Y) \Phi^*_{l,Q_E}(Y) \Phi_{l,Q_E}(Y')=\delta(Y-Y'),
\ee
or symbolically by also omitting the $l$-dependence,
\begin{eqnarray}
&&\int dY G(Y)\;|Y\rangle \langle Y|=1,\quad
\int dE \;|E \rangle \langle E|=1\\
&&G(Y)\langle Y|Y'\rangle=\delta(Y-Y'), \quad \langle
E|E'\rangle=\delta(E-E').
\end{eqnarray}
where $Y:=(u,v,x)$.

  Moreover, from the orthonormal and complete conditions \eq{or1},
\eq{co1}, we can evaluate the trace for any operator ${\cal A}$ as
following (we omit the $l$-dependence for simplicity)
\begin{eqnarray}
\Tr \;{\cal A}&=&\int dY G(Y) \langle Y|{\cal A}|Y\rangle \nonumber\\
&=&\int dY dE G(Y)|\langle Y|E\rangle|^2 \langle E|{\cal
A}|E\rangle. \label{TrcalX}
\end{eqnarray}

Then, the Gaussian path integral for the effective action can be
evaluated as following
\be\label{DetGD}
e^{\Tr \log i \Delta_l}=\int DX e^{i\int d^4x  \;G_l(u,v,x)
X\Delta_l X}=\prod_{E_l}  {i \over \sqrt{E_l}}.
\ee

The introduction of the non-trivial measure $G_l$ in defining the inner product is the key step in dealing with the quantum field theory in warped plane-wave space in contrast to the one in the unwarped case.

\bigskip

  Now we would like to exactly solve the eigenvalue equation \eq{eigen1}
for a background that is as general as possible. This is the case if we put the
following ansatz
\begin{eqnarray*}
F(u):=a^2(u)b^2(u)=F_0+F_1(u), \qquad h(u,x^i)=h_0+h_1(u)+h_2(x^i).
\end{eqnarray*}
 Even though the background metric and scalar are quite arbitrary,
we can still solve the equation of motion exactly due to the
peculiar nature of the plane-wave metric rendering the first order
equation of motion in $u$.  Then the solution is
\begin{eqnarray}\label{noBobo}
\Phi_{l,E}(u,v,x)&=&\frac{1}{(\sqrt{2\pi})^2\sqrt{G_l}}\;
e^{[i(uq+pv)+\frac{i}{2p}\int^u (F_1(s)(1-\delta_{a,3})-p^2h_1(s))
ds]}K_p(x^i)
\end{eqnarray}
with $G_l$ given in \eq{Gs1}, and $K_p(x^i)$ satisfies following
eigenvalue equation
\begin{eqnarray}
[\partial_i^2+p^2h_2(x^i)]K_p(x^i)=-k^2 K_p(x^i) \label{eigen3}
\end{eqnarray}
The corresponding eigenvalue is
\begin{eqnarray}\label{spectrum1}
E&=&2qp+h_0p^2-k^2-F_0(1-\delta_{a,3}).
\end{eqnarray}
Note that the eigenvalue $k^2$ and eigenfunction $K_p$ are implicit
functions of $p$. The eigenfunctions $\Phi_{l,E}$'s satisfy the
orthonormal and complete conditions \eq{or1}, \eq{co1} if $K_p$ is a
solution of \eq{eigen3} which satisfies the complete condition.

It is interesting to note that the energy spectrum \eq{spectrum1} is
independent of $l$, and the background dependence of the
eigenfunction only appear as a pure phase and yield no physical
effect on the energy spectrum. This is due to the peculiar feature
of the plane-wave-like background such that the equation of motion
is of first order in $u$. Moreover, despite the nontrivial
time-dependent scale factor and background vev, we see that the
solution is still of plane-wave type. This implies that there is no
particle production which is induced by the Bogolubov coefficient,  contrary to the usual expectation for the quantum cosmological particle production. This fact is also due to
the peculiar feature of the plane-wave-like metric.

  From \eq{DetGD} and \eq{TrcalX}  we find that
 \begin{eqnarray}
 tr_{SU(2)} \Tr \log i\Delta_l &=& \int d^4x\;
\int d^4p
\frac{1}{(2\pi)^2}|K_p(x^i)|^2 \nonumber\\
&&\left[2\log i(2pq-k^2+h_0p^2-F_0)+\log i(2pq-k^2+h_0p^2)\right]
\nonumber\\
&=&-\int d^4x\; \int d^4p
\frac{1}{(2\pi)^2}|K_p(x^i)|^2\nonumber\\
&&\hspace{5mm}\int^{\infty}_{0}\frac{ds}{s}e^{-is(2pq-k^2+h_0p^2)}\left[2e^{isF_0}+1\right]
\end{eqnarray}
where  we have introduced the Schwinger's representation as in the
usual heat-kernel method in the last equality. Also $\int d^4p$ is
the abbreviated form of $\int dqdpd^2k$.

 For the last term in \eq{veb2} we have
\begin{eqnarray*}
&&\hspace{-5mm} tr_{SU(2)}\delta_{a,3}\Tr\int d^4x
\frac{1}{2}e^{-\phi}a^2b'^2
\left(\overrightarrow{\partial_v}\frac{1}{\Delta_3}\overleftarrow{\partial_v}\right) \\
&&\hspace{-10mm}= \int d^4x\; \int d^4p
\frac{1}{(2\pi)^2}|K_p(x^i)|^2 \frac{1}{2}b'^2p^2
\Bigg(
\frac{1}{2pq-k^2+h_0p^2}
\Bigg)
\\
&&\hspace{-10mm}=-i\int d^4x\; \int d^4p
\frac{1}{(2\pi)^2}|K_p(x^i)|^2
\frac{1}{2}b'^2p^2 \int^{\infty}_{0}ds\;e^{-is(2pq-k^2+h_0p^2)}\\
\end{eqnarray*}

   Summing all up,  the 1-loop bosonic vacuum energy becomes
\begin{eqnarray*}
\hspace{-5mm}-i\int d^4x\; \int d^4p
\frac{1}{(2\pi)^2}|K_p(x^i)|^2\int^{\infty}_{0}ds\;
e^{-is(2pq-k^2+h_0p^2)} \Bigg[
\frac{1}{2}b'^2p^2-\frac{4}{s}\left(2e^{isF_0}+1\right) \Bigg].
\end{eqnarray*}
There is an interesting simplification here. The $q$-integration gives
us the delta function $2\pi\delta(2sp)=\pi\delta(p)/s$. Thus after
$p$-integration we arrive
\begin{eqnarray}
\Lambda_b&=&4i\pi\int d^4x\; \int d^2k \frac{1}{(2\pi)^2}|K_{p
\rightarrow 0}(x^i)|^2 \int^{\infty}_{0}\frac{ds}{s^2}\;
e^{isk^2(0)}\left(2e^{isF_0}+1\right)\label{lambb}
\\\nonumber
&:=&\int d^4x\sqrt{-g}\;\lambda_b.
\end{eqnarray}
We then find that the effective vacuum energy density $\lambda_b$(
or the ``cosmological constant") is time-dependent. Moreover, the
vacuum energy depends on the background scalar vev $b(u)$ through
$F_0$ but not $F_1(u)$. Therefore, the scalar background $b(u)$ contributes,
only when it behaves $b\sim a^{-1}$. In this case, the contraction
of the separation between two D3-branes is counterbalanced by the
expansion of the Universe so that it is not red-shifted or
blue-shifted, then it contributes to the vacuum energy.

\subsection{Fermionic sector}

  The eigenvalue equation for evaluating the fermionic effective action is
\be\label{dirac1}
i\;\Gamma^0\;\Dslash \Psi=\lambda\Psi
\ee
where the Dirac operator $\;\Dslash$ is given in \eq{Dslash1}.

 Instead of solving \eq{dirac1}, we will solve the second order differential equation
\begin{eqnarray}
(\Delta_4+V-\lambda^2)\Psi=0
\label{dirac2}
\end{eqnarray}
with
\begin{eqnarray}
\Delta_4&=&\square
+((\log a)'-\phi')\partial_v-F (1-\delta_{a,3}),\\
V&=&3(\log a)'\partial_v
-\Gamma^+\Gamma^i \frac{1}{2}\partial_{x_i} h\partial_v
+\Gamma^+\Gamma^m \partial_u(a{\cal B}). \label{Vfdet}
\end{eqnarray}
As usual equation \eq{dirac2} can be easily derived by squaring the
Dirac equation \eq{dirac1} with the help of \eq{Dslash1}. Here we
regard $V$ as a perturbation.

The eigenfunction of $\Delta_4$ is
\begin{eqnarray*}
\Psi_{E}(u,v,x)&=&\frac{1}{(\sqrt{2\pi})^2\sqrt{G_4}}\;
e^{[i(uq+pv)+\frac{i}{2p}\int^u (F_1(s)(1-\delta_{a,3})-p^2h_1(s))
ds]}K_p(x^i)
\end{eqnarray*}
with $G_4=e^{-\phi}/a$, and  the eigenvalue
\be
E=2pq-k^2+p^2h_0-F_0(1-\delta_{a,3})
\ee
which is the same as \eq{spectrum1}. However, the full fermionic spectrum is different from the bosonic one \eq{spectrum1} by the additional contribution from \eq{Vfdet}, which is pure imaginary as shown below.

In terms of the 4-dimensional scalar determinant, the one for the
Dirac operator is
\be
\det i a\; \Dslash= [\det i^2 (\Delta_4+V)]^4
\ee
by using the fact that the  ${\cal N}=4$ SYM contains 4 Dirac
fermions. Therefore, the fermionic effective vacuum energy is
nothing but
\be
\Lambda_f=4i\; tr_{SU(2)} \Tr \log i(\Delta_4+V).
\ee
 Though the perturbation $V$ looks complicated, however, from the fact
\be
 \Tr(\Gamma^+\Gamma^i)^n=0,\qquad \Tr(\Gamma^+\Gamma^9)^n=0 \qquad  \mbox{for integer}\; n\ge 1,
\ee
we see that only the first term  in $V$ can survive the perturbation
series. Therefore, the ``eigenvalue" of $V$ is pure imaginary
$3i(\log a(u))'p$ so that the on-shell condition implies that $p=0$
for nonzero $(\log a(u))'$. This implies that the on-shell fermion
in null-like cosmological background \eq{metric4} has zero light
cone momentum. It is interesting to see that there is no such a
constraint for bosons.

With all the above, the fermionic vacuum energy can be evaluated
explicitly
\begin{eqnarray*}
\Lambda_f&=&4i\; tr_{SU(2)} \Tr \log i(\Delta_4+3(\log a(u))'\partial_v)\\
&=&-4 i \int d^4x\; \int d^4p \frac{1}{(2\pi)^2}|K_p(x^i)|^2
\int^{\infty}_{0}\frac{ds}{s}e^{-is(2pq-k^2+h_0p^2+3i(\log
a)'p)}\left[2e^{isF_0}+1\right]
\\
&:=& \int d^4x\sqrt{-g}\;\lambda_f.
\end{eqnarray*}
After $q$-integration to yield the delta function
$2\pi\delta(2sp)=\pi\delta(p)/s$, we find that the resulting
$\Lambda_f$ is exactly canceled by the bosonic vacuum energy
\eq{lambb} though the fermionic and bosonic spectrum are matched
only in the real part. This cancellation is due to the manifest
supersymmetries which are broken to $1/4$ by the background even though the fermionic and bosonic spectrum are different by an imaginary term.

\section{Examples}
In previous section we have shown that the 1-loop
vacuum energy of the dual SYM vanishes for the generic background
\eq{generic1} and \eq{b9} as long as the eigenfunction $K_p(x^i)$ is
solved. Here we would like to carry out the evaluation for some
specific cases for pedagogical consideration.

 Recall that the Einstein equation which constrains the scale factor $a(u)$, dilaton $\phi(u)$ and
the plane-wave profile $h(u,x^i)$ is
\begin{eqnarray*}
2((\partial_u\ln a)^2-\partial_u^2\ln a)
-\frac{1}{2}\vec{\nabla}^2h
=\frac{1}{2}\partial_{u}\phi\partial_u\phi.
\end{eqnarray*}
Formal solution of $\phi(u)$ and $h(u,x^i)$ for a given $a(u)$ is
\begin{eqnarray*}
\phi(u)&=&\pm 2\int^u  \sqrt{
((\partial_s\ln a(s))^2-\partial_s^2\ln a(s))-\tau^2(s)}\;ds,\\
\\
h(u,r,\theta)&=&4\tau(u)^2r^2+\tilde{h}(r,\theta),\\
\tilde{h}(u,r,\theta)&=&A_1(u)\log r+A_2(u)\\
&&+\sum_{n=1}^{\infty}(A_3(u)r^{n(A_5(u)+A_6(u))}+A_4(u)r^{-n(A_5(u)+A_{6}(4))})
(A_5(u)e^{in\theta}+A_6(u)e^{-in\theta}).
\end{eqnarray*}
where $(r,\theta)$ is polar coordinate of $x_2,x_3$ plane,
$\tilde{h}$ satisfies 2-dimensional Laplace equation $\nabla^2
\tilde{h}=0$, $\tau(u)$ and $\{A_1(u), .... A_6(u)\}$ are functions
of $u$. We will give two examples in which the eigenvalue equation
can be solvable exactly.

\subsection{Example I: Time-dependent Weyl-flat space background}
If the spacetime is Weyl-flat for any scale factor $a(u)$, we then
have
\begin{eqnarray*}
h(u,r,\theta)&=&h_0 = const,\\
 \phi(u)&=&\pm 2\int^u  \sqrt{
((\partial_s\ln a(s))^2-\partial_s^2\ln a(s))}\;ds.
\end{eqnarray*}
In this case $K_p(x^i)=e^{k\cdot x}$ solves \eq{eigen3}, and the
eigenfunction of $\Delta_l$ is
\begin{eqnarray*}
\Phi_{l,E}&=&\frac{1}{\sqrt{(2\pi)^4}}\frac{1}{\sqrt{G_l}}e^{i(qu+pv+k\cdot x)+\frac{i}{2p}
\int^u(F_1(s)(1-\delta_{a,3})-p^2h_1(s))ds},\\
E&=&2qp-k^2+p^2h_0-F_0(1-\delta_{a,3}).
\end{eqnarray*}
These basis are orthogonal to each other and complete.

The 1-loop vacuum energy can be easily obtained as
\begin{eqnarray*}
\Lambda&=&\frac{1}{(2\pi)^4}\int dx^4 \int d^4p \Bigg[
\frac{1}{2}b'^2
p^2\left(\frac{1}{2pq-k^2+h_0p^2}\right)\\
&&-2\cdot 4i \log i(2pq-k^2+h_0p^2-F_0) +2\cdot 4i\log i(2qp-k^2+h_0p^2-F_0+3i(\log a)'p)\\
&&-4i \log i(2pq-k^2-h_0p^2) +4i\log i(2qp-k^2-h_0p^2+3i(\log a)'p)
\Bigg]
\end{eqnarray*}
which vanishes as explained in the previous section.

On the other hand, bosonic YM theory has following 1-loop potential:
\begin{eqnarray*}
\Lambda_b&=&\frac{1}{4\pi^2} \int dx^4
\; \int_{0}^{\infty}\frac{ds}{s^3}
(2e^{isF_0}+1).
\end{eqnarray*}

\subsection{Example I{}I: Time-dependent Weyl-plane-wave background}
For the Weyl-plane-wave spacetime with arbitrary scale factor
$a(u)$, we have
\begin{eqnarray*}
h(x^i)&=&-4\tau^2(x_2^2+x_3^2)+h_0,\\
\phi(u)&=&\pm 2\int^u  \sqrt{ ((\partial_s\ln
a(s))^2-\partial_s^2\ln a(s))+\tau^2}\;ds, \qquad \lambda=const.
\end{eqnarray*}
In this case the eigenvalue equation solved by $K_p(x_i)$ is
\begin{eqnarray*}
\sum_i\left(-\partial_i^2+4\tau^2p^2x_i^2\right)K_p=k^2K_p.
\end{eqnarray*}
Thus $K_p(x_i)$ is nothing but the wave function of harmonic
oscillator in two dimensional plane:
\begin{eqnarray*}
K_{p,n,m}(x_2,x_3)&=& \sqrt{\frac{\sqrt{2p\tau}}{2^nn!\sqrt{2\pi}}}
\sqrt{\frac{\sqrt{2p\tau}}{2^mm!\sqrt{2\pi}}}
H_n(\sqrt{2p\tau}x_2)H_m(\sqrt{2p\tau}x_3)e^{-p\tau(x_2^2+x_3^2)}
\end{eqnarray*}
with the eigenvalue $k^2=4p\tau(n+m+1)$.
$H_n$ is the Hermite polynomial defined as
\begin{eqnarray*}
H_n(x)=(-1)^ne^{x^2}\frac{d^n}{dx^n}e^{-x^2}.
\end{eqnarray*}
It satisfies the orthogonal and completeness conditions
\begin{eqnarray*}
\int^{\infty}_{\infty}dx \left(\sqrt{\frac{1}{2^nn!\sqrt{2\pi}}}H^{*}_n(x)e^{-\frac{x^2}{2}}\right)
\left(\sqrt{\frac{1}{2^mm!\sqrt{2\pi}}}H_m(x)e^{-\frac{x^2}{2}}\right)
&=&\delta_{n,m},\\
\\
\sum_{n=0}^{\infty} \left(\sqrt{\frac{1}{2^nn!\sqrt{2\pi}}}H^{*}_n(x)e^{-\frac{x^2}{2}}\right)
\left(\sqrt{\frac{1}{2^nn!\sqrt{2\pi}}}H_n(y)e^{-\frac{y^2}{2}}\right)
&=&\delta(x-y).
\end{eqnarray*}
Then the eigenfunctions and eigenvalues of $\Delta_l$ are
\begin{eqnarray*}
\Phi_{l,E}&=&\frac{1}{\sqrt{(2\pi)^2}}\sqrt{\frac{\sqrt{2p\tau}}
{2^nn!\sqrt{2\pi}}} \sqrt{\frac{\sqrt{2p\tau}}{2^mm!\sqrt{2\pi}}}
\sqrt{\frac{1}{G_l}}\\
&&\;\times\;e^{i(qu+pv)+\frac{i}{2p}\int^u(F_1(s)(1-\delta_{a,3})-p^2h_1(s))ds}
H_n(\sqrt{2p\tau}x_2)H_m(\sqrt{2p\tau}x_3)e^{-p\tau(x_2^2+x_3^2)},\\
\\
E_l&=&2pq-4p\tau(n+m+1)+p^2h_0-F_0(1-\delta_{a,3})
\end{eqnarray*}
with positive integer $m$ and $n$. The 1-loop effective vacuum energy is
\begin{eqnarray*}
\Lambda&=&
\frac{1}{(2\pi)^3}\int dx^4 \;e^{-(x_2^2+x_3^2) }
\int dqdp\sum_{n,m=0}^{\infty} \frac{1}{2^nn!}\frac{1}{2^mm!}
|H_n(x_2)|^2|H_m(x_3)|^2 \\
&& \hspace{-10mm} \Bigg[ \frac{1}{2}b'^2 p^2 \left(
\frac{1}{2pq-4p\tau(n+m+1)+p^2h_0}\right)\\
&&\hspace{-8mm} -2\cdot 4i \log
i(2pq-4p\tau(n+m+1)+h_0p^2-F_0)\\
&&\hspace{-8mm}+2\cdot 4i\log i\left(2qp-4p\tau(n+m+1)+h_0p^2-F_0+3i(\log a)'p\right)\\
&&\hspace{-8mm} -4i \log i(2pq-4p\tau(n+m+1)-h_0p^2) +4i\log
i\left(2qp-4p\tau(n+m+1)-h_0p^2+3i(\log a)'p\right) \Bigg].
\end{eqnarray*}
It vanishes by boson-fermion cancellation. On the other hand, the
bosonic part of the vacuum energy is
\begin{eqnarray*}
\Lambda_b=-\Lambda_f =\frac{1}{\pi}\int dx^4\;\delta^2(0)
\int_{0}^{\infty}\frac{ds}{s} ( 2e^{isF_0}+1).
\end{eqnarray*}

\section{Comments on particle production and UV structure}

   The quantum field theories discussed in this paper are very different from the conventional ones because of their null-like time-dependence in both coupling constant and scalar vevs on a cosmological background. It is then interesting to discuss some issues due to these peculiarities.

\begin{itemize} 

\item    The first issue we would like to discuss is about the particle production. 
This issue has been discussed in \cite{conformal}, here we would like to point 
out some more subtleties.  In our case, the cosmological particle production is inhibited because the Bogolubov coefficients are trivial 
as implied by the solution \eq{noBobo}, which can be defined globally in time direction. We may still expect particle production due to the self-interaction because the lack of Poincare invariance allows energy and momentum to emerge from the vacuum.  The particles will be produced by some time-dependent interaction sourced by $J$,  which comes from the time-dependent factors of the background and coupling. This can be characterized by the squeeze state 
\be\label{squeeze}
|s\rangle= \exp(-i \int J(t) {\cal L}_{int}) |0\rangle.
\ee
Explicitly we can take ${\cal L}_{int}=\lambda \varphi^3$, and expand the second quantized field $\varphi$ in terms of the positive-energy modes.

  However, as pointed out in \cite{conformal}, there will be no particle production in the quantum field theories on the null-like time-dependent background even though there is a time-dependent source. This is mainly due to the fact that the light-cone energy $q$ is related to the light-cone momentum $p$ and the spatial momenta $k_i$ by \eq{spectrum1}, such that on shell we have 
\be
q={k^2+F_0(1-\delta_{a,3})-h_0p^2 \over 2p}.
\ee
In \cite{conformal} the authors consider the case with $F_0=h_0=0$ so that the positive-energy condition $q\ge 0$ requires $p>0$. On the other hand, the null-like background considered here preserves the translational invariance. It will then impose a delta function for the null-like momentum conservation $\delta(p_1+p_2+p_3)$ in \eq{squeeze}. This condition is, however, incompatible with $p_i>0$ so that there will be no particle production. The above argument holds true for $h_0\le 0$ case (note that $F_0\ge 0$ is given by definition).

   We would like to point out a subtlety here: for the background such that $h_0>0$, it is then possible to have $q\ge 0$ even for $p< 0$, as long as $h_0p^2>k^2+F_0(1-\delta_{a,3})$. Therefore, the no-go of particle production in $h_0\le 0$ case is lifted, and then the particle production induced by time-dependent source is possible. It is interesting to see the relation between the quantum stability and the sign of $h_0$ though its physical origin is not clear and deserves further study. 
   
   The above discussion is for bosonic case only. For fermion we recall that there is an additional on-shell condition $p=0$, which gives the on-shell condition on the transverse momentum: 
\be   
   k^2+F_0(1-\delta_{a,3})=0
\ee   
Since $F_0$ is positive by definition, this yields no fermion production.

\item The second issue we would like to comment on is the possible new UV structure of the theory induced by the time dependence of the background and coupling. Combining the time-dependent factors of background and coupling, we can realize the SYM theory as the one in flat space but with the time-dependent effective coupling. This theory is novel from the point of view of the conventional quantum field theory. For example, the interaction vertices in the momentum space may not have the delta function conserving the energy as in the usual QFT. More explicitly, we could obtain interaction
vertices by eigenfunction expansion of the fields.
Every interaction vertex will then have the following form
\be
\delta(\sum_i
p_i)\delta^2(\sum_i k_i)\alpha(\sum_iq_i)
 \qquad \mbox{$i$ labels the lines attached to the vertex},
\ee
and $\alpha(\sum_iq_i)$ is the Fourier transformation of $u$-dependent
effective coupling, and can be further Taylor (i.e., derivative) expanded to 
yield new vertices. In general $\alpha(\sum_iq_i)$ dose not yield the delta 
function for energy conservation, which is, however, expected because 
there is no global time-like Killing vector. 

   Now it is clear that all the novelty of the theory is summarized in the 
Fourier transform of the time-dependent effective coupling $\alpha(q)$. 
If this function is UV finite, then the UV structure of the theory will reduce 
to the one for the ordinary SYM.  In this case we may investigate the loop 
effect by Taylor expansion of the time-dependence of the effective coupling, 
and this corresponds to the derivative expansion of $\frac{\partial}{\partial q}$ 
in the momentum space. In this way many (possibly infinite) new interaction 
vertices will appear.  Since it is well known that the gauge theory has 
only log divergence in flat space, therefore the higher derivative interactions 
will contribute no more than the log divergence in the loop effect, 
and then will be suppressed in the UV limit. This implies that the UV structure 
of these time-dependent SYM's will be the same as the ordinary one. Especially, 
the beta function for the gauge coupling will be zero as in the ordinary SYM. 
This then implies that the theory is scale-invariant. 
   
   On the other hand, if $\alpha(q)$ is not UV-finite, then we cannot employ the derivative expansion to calculate the loop effect. Instead we may need to do the calculation in the coordinate space with subtle renormalization scheme. In this case we may see the singular structure in the higher derivative expansion, which may lead to singular S-matrix in a way similar to the time-dependent cases discussed in \cite{LMS}. In contrast, these non-trivial UV behaviors have not appeared in 1-loop vacuum energy since the 1-loop determinant is independent of the effective coupling. However, we expect they will appear in vacuum energy beyond 1-loop or will appear in 1-loop corrections of correlation functions.
We hope to investigate this issue in the near future.

\end{itemize}

\section{Conclusion}
Our calculation of 1-loop effective potential of dual SYM, on
4-dimensional null-like time-dependent backgrounds preserving
1/4-supersymmetry, yields zero 1-loop correction for quite generic
situation. Moreover, we argue that the UV structure is the same as that in the 
ordinary SYM as long as the Fourier transform of the time-dependent effective coupling
is UV-finite. Altogether, this suggests the existence of
non-renormalization theorem for such a time-dependent theory, and
gives direct evidence that the dual SYM is scale invariant, which was
conjectured in \cite{conformal} in order to obtain the power-law
behavior of the SYM correlators, and partly verifies the holographic principle
discussed earlier in \cite{ch}. 

  Besides the implication on the SYM side, our result also yields new understanding about the AdS string.  From the S-duality of the AdS/CFT correspondence, our result implies the bulk quantum gravity near the big bang could behave regularly as suggested by our 1-loop result, at least for the UV-finite effective coupling. Moreover, our result also gives the evidence for the 1-loop triviality of the open
string partition function in the dual time-dependent AdS space. This
generalizes the result for the unwarped plane-wave as directly evaluated in \cite{hikida},
and it is non-trivial in the sense that the perturbative  string
theory in AdS space is hard to calculate. It will be interesting to
explore more direct connections implied by the holographic principle
for such a null-like time-dependent background.

\bigskip

\acknowledgments We would like to thank Pei-Ming Ho, Ta-Sheng Tai
and Wen-Yu Wen for discussions. We also thank Ming-Che Chang for carefully reading our draft and give advices. This work is supported by Taiwan's
NSC grant 94-2112-M-003-014 and 95-2811-M-003-007.

\begin{appendix}
\section*{Appendix: Gaussian Integrations over bosonic
fluctuations}

   In the following we will perform the Gaussian integrations to obtain the effective action for the background configuration $B^m$
with $u$ dependence.

   Start from the action \eq{SA1} and use the basic formula
\begin{eqnarray*}
\frac{1}{2}(aAa+J^{\dagger}a+aJ+B)= \frac{1}{2}\left(a+J^{\dagger}\frac{1}{A}\right)A\left(a+\frac{1}{A}J\right)
-\frac{1}{2}J^{\dagger}\frac{1}{A}J+\frac{1}{2}B,
\end{eqnarray*}
we first integrate over $A_i$ and it yields the effective action
\begin{eqnarray}
&&i\Tr \log i\Delta_1+i\Tr\int d^4x\frac{1}{2}\Bigg[
-(e^{-\phi}\Delta_1A_{u})A_v\nonumber
\\
&&-A_v(e^{-\phi}\Delta_1A_u) +A_v
e^{-\phi}(H\frac{1}{\Delta_1}H-h\Delta_2)A_v -2i{\cal J}A_v \Bigg].
\end{eqnarray}
$2i{\cal J}A_v$ is linear term of $A_v$. The explicit  form is
\begin{eqnarray*}
{\cal J}=2[\partial_uB^m, X^m]+\partial_u(e^{-\phi}a^2)({\cal B}X).
\end{eqnarray*}

Note that we have written $i\Tr \log i\Delta_1$, not $i\Tr\log
ie^{-\phi}\Delta_1$, as a result of Gaussian integral. The reason is
that here we expand the fields $A_{\mu}$ with the eigenfunctions of
$\Delta_1$, which satisfy
\begin{eqnarray}
&&\qquad \qquad \Delta_1 \Phi_{1,E}=E\Phi_{1,E}, \\
&& \int d^4x e^{-\phi}\Phi^*_{1,E}(u,v,x) \Phi_{1,E'}(u,v,x)=\delta^{(4)}(E-E').
\end{eqnarray}
We see that the measure factor for the orthonormality of the
eigen-base is $G_1:=e^{-\phi}$. Similar re-definition of the inner
product with the corresponding measure factor will also be done for
the Gaussian integrals over $A_v$, $A_u$ and $X$ in the following.

 Next we integrate the above over $A_v$, the effective action becomes
\begin{eqnarray}
&&i\Tr \log i\Delta_1+\frac{1}{2}i\Tr \log
i(H\frac{1}{\Delta_1}H-h\Delta_2)\nonumber\\
&&+i\Tr\int d^4x\frac{1}{2}\Bigg[
-(e^{-\phi}\Delta_1A_{u}+i{\cal J})\nonumber\\
&& \frac{e^{\phi}}{H\frac{1}{\Delta_1}H-h\Delta_2 }
(e^{-\phi}\Delta_1A_{u}+i{\cal J}) \Bigg].
\end{eqnarray}

   Then we integrate the above over $A_u$, and combine the result
with $S_X$ in \eq{SX1} to yield the effective action
\begin{eqnarray}
&&i\Tr \log i\Delta_1+\frac{1}{2}i\Tr \log i(H\frac{1}{\Delta_1}H-h\Delta_2)
\nonumber\\
&&+\frac{1}{2}i\Tr \log \left( \Delta_1
\frac{-i}{H\frac{1}{\Delta_1}H-h\Delta_2}\Delta_1\right) +i\Tr\int
d^4x\frac{e^{-\phi}}{2}\Bigg[ -2a^2\partial_uB^m\partial_vX^m
\nonumber \\
&&+a^{2}\partial_{u}X\partial_{v}X+a^{2}\partial_{v}X\partial_{u}X
+a^2h\partial_{v}X\partial_{v}X -a^2\partial_{i}X\partial_{i}X
-a^4X^{m}{\cal B}^2X^m \Bigg].
\end{eqnarray}

  Finally, we integrate the above over $X$ to yield the bosonic effective action by also adding the ghost contribution
\begin{eqnarray}\label{Sbeff1}
S_{b,eff}&=&i\Tr \log i\Delta_1+\frac{1}{2}i\log i(H\frac{1}{\Delta_1}H-h\Delta_2)
\nonumber\\
&&+\frac{1}{2}i\Tr \log\left( \Delta_1
\frac{-i}{H\frac{1}{\Delta_1}H-h\Delta_2}\Delta_1\right)
-i\Tr \log \Delta_0\nonumber\\
&&+3i\Tr \log i\Delta_3 -\Tr\int d^4x\frac{1}{2}e^{-\phi}a^2\Bigg[
(\partial_u B)\overrightarrow{\partial_v} \frac{1}{\Delta_3}
\overleftarrow{\partial_v}(\partial_u B) \Bigg].
\end{eqnarray}
Note that the measure factor for the inner product of the eigen-base
of $\Delta_3$ is $G_3:=e^{-\phi}a^2$.

In the above, we have defined
\begin{eqnarray}
\square&:=&-2\partial_{u}\partial_{v}-h\partial_{v}\partial_{v}
+\partial_{i}\partial_{i} .
\\
\Delta_0&:=& \square-a^2{\cal B}^2\\
\Delta_1&:=& \square +\phi'\partial_v-a^2{\cal B}^2\\
\Delta_2&:=& \square -2\frac{h'}{h}\partial_v+\frac{\nabla_ih}{h}\partial_i-a^2{\cal B}^2\\
\Delta_3&:=&\square +\left(\phi'-2\frac{a'}{a}\right)\partial_v
-a^2{\cal
B}^2\\
H&:=&-\phi'\partial_i-\nabla_ih\partial_v
\end{eqnarray}
Note that for the configuration $B^m$ given in \eq{b9},  \eq{Sbeff1}
will reduce to \eq{veb1}.

\end{appendix}

%%%%%%%%%%%%%%%%%%%%%%%%%%%%%%%%%%%%%%%%%%%%%%%%%%%%%%%%%%%%%%%%%%%%%%%%%%%%

\end{document}